\documentclass[12pt,a4paper]{article}
\usepackage{graphicx,epsfig}
\usepackage{color}
\usepackage{amsmath}
\usepackage{amsfonts}
\usepackage{amssymb}
\usepackage{cite}
\usepackage{ifpdf}
\newcommand{\be}{\begin{equation}}
\newcommand{\ee}{\end{equation}}
\newcommand{\bea}{\begin{eqnarray}}
\newcommand{\eea}{\end{eqnarray}}
\newcommand{\ba}{\begin{array}}
\newcommand{\ea}{\end{array}}

\newcommand{\cO}{{\cal O}}

\newcommand{\no}{\nonumber}

\newcommand{\lsim}{\stackrel{<}{_\sim}}
\newcommand{\gsim}{\stackrel{>}{_\sim}}

\definecolor{Red}{rgb}{1.,0.,0.}

\begin{document}

\begin{center}
{\Large \bf ElectroWeak Precision Observables at One-Loop in Higgsless models}\vskip 1.0cm
{\large  Oscar Cat\`a$^a$ and Jernej~F.~Kamenik$^{b}$} \\[0.5 cm]
{\it $^a$~~Ludwig-Maximilians-Universit\"at M\"unchen, Fakult\"at f\"ur Physik,\\
Arnold Sommerfeld Center for Theoretical Physics, 
D--80333 M\"unchen, Germany}\\
{\it $^b$~J.~Stefan Institute, Jamova 39, P.O.~Box 3000, 1001 Ljubljana, 
Slovenia}\\ [1.0 cm]
{\bf Abstract}
\begin{quote}
We study the viability of generic Higgsless models at low energies when compliance with EWPO and unitarity constraints up to the TeV scale are imposed. Our analysis shows that consistency with $ S$ and $ T$ can be achieved at the one-loop level even with a single light vector state, $m_V\lsim 500$ GeV. However, this scenario turns out to be strongly disfavored when direct resonance searches at the Tevatron are also taken into account. We show that a fully consistent picture can be obtained if an axial state is introduced. Interestingly, $m_V$ is still predicted to be light (below $1$~TeV) while typical values of $m_A$ span over the window $1.2~m_V\leq m_A\leq 1.4~m_V$. Our results for the vector channel are rather robust and well within the reach of present-day colliders, while the axial channel is more loosely constrained. 
\end{quote}
\end{center}
\vskip 1.0cm
%%%%%%%%%%%%%%%%%%%%%%%%%%%%%%%%%%%%%%%%%%%%%%%%%%%%
\section{Introduction} 
\begin{figure}[t]
\begin{center}
\includegraphics[width=8.cm]{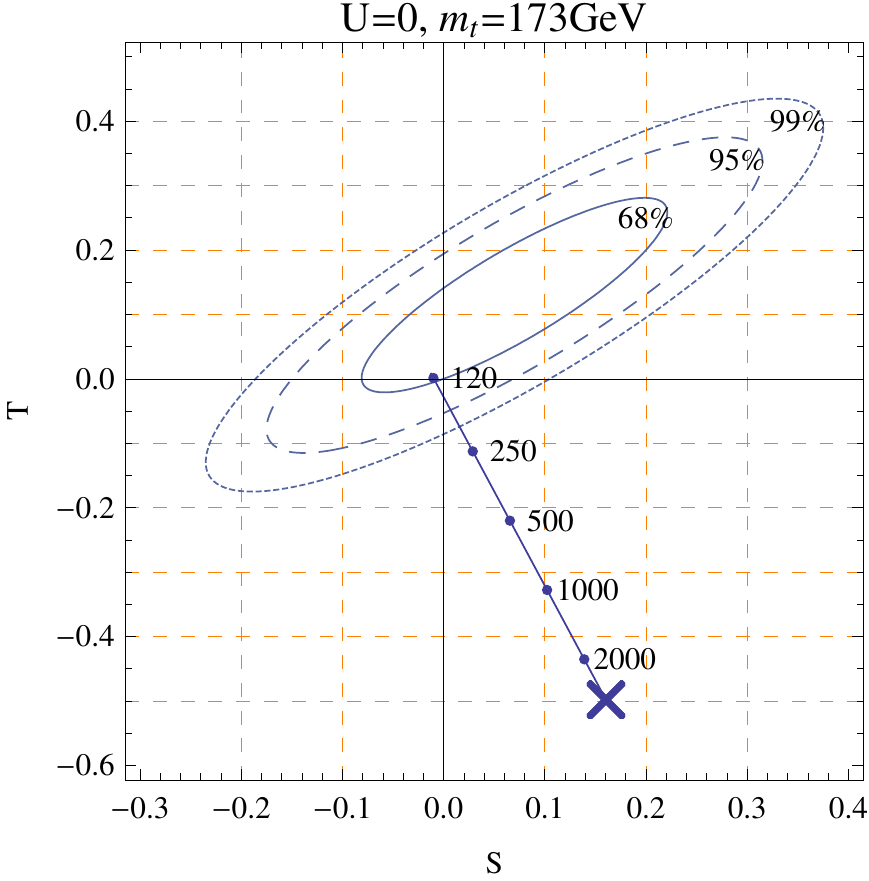}
\end{center}
\caption{\label{fig:1} {\it{{\footnotesize{IR logarithmic contributions to $S$ and $T$ for a  cut-off scale $\Lambda$ in the range $115 - 3000$~GeV. The experimentally allowed region and the SM reference point at which all $STU$ parameters vanish is chosen as in Ref.\cite{:2005ema} (in particular with $m_t = 175 $GeV and $m_H = 150 $GeV). The constraint $U = 0$ is always imposed. The axes are scaled to $S = 4 \sin^2\theta_W \hat S / \alpha \approx 119 \hat S$ and $T = \hat T/\alpha \approx 129 \hat T$.}}}}}
\end{figure}

%%%%%%%%%%%%%%%%%%%%%%%%%%%%%%%%%%%%%%%%%%%%%%%%%%%%
Up to the present date, the physical mechanism underlying electroweak symmetry breaking (EWSB) still remains unknown. Its simplest description in terms of a fundamental standard model (SM) scalar Higgs doublet is certainly appealing but not fully satisfactory from a theoretical standpoint. A single Higgs doublet makes the electroweak interactions renormalizable and, if light enough ($m_H \sim 100$~GeV), all the electroweak precision tests (EWPT) can be successfully accounted for~\cite{:2005ema} (see Fig.~\ref{fig:1}). However, $m_H$ is unstable under quantum corrections: at short distances it diverges quadratically and thus makes the SM description problematic. One is therefore led to consider alternative mechanisms of EWSB, either with Higgs bosons ({\it{e.g.}} supersymmetry) or without them (Higgsless theories).

In Higgsless models the Higgs boson is often replaced by one or more spin-1 bosons. The theory is then no longer renormalizable, but perturbative unitarity can still be preserved up to a few TeV, one order of magnitude above the Fermi scale, $v = (\sqrt 2 G_F )^{-1/2} \simeq 246$~GeV, so that meaningful comparisons with current experiments can be made. The ultraviolet completion of the theory will depend on the underlying Higgsless model, be it technicolor, deconstructed models with hidden-gauge symmetry or holographic realizations, but at low energies they all should be described in terms of a small subset of effective operators with its associated couplings. Obviously, depending on the specific Higgsless model one considers there will be different relations amongst the couplings. In this paper we will instead pursue a more phenomenological approach, following~\cite{Barbieri:2008cc}, and consider a generic Higgsless model whose parameters will be constrained only by unitarity and electroweak precision observables ($S$ and $T$) up to the one-loop level. The main advantage of this approach is that one can test the viability of Higgsless models based on spin-1 bosons as model-independently as possible (and potentially rule out some of the existing models). Additionally, one can extract information on the spin-1 bosons which can then be tested at colliders.

Oblique corrections to electroweak precision observables (EWPO) can be conveniently expressed in terms of the SM gauge boson vacuum polarization correlators, defined as
\begin{equation}
\mathcal L_{vac-pol} = -\frac{1}{2} W_\mu^3 \Pi_{33}^{\mu\nu}(q^2) W_\nu^3 - \frac{1}{2} B_\mu \Pi_{00}^{\mu\nu}(q^2) B_\nu - W_\mu^3 \Pi_{30}^{\mu\nu}(q^2) B_\nu - W_\mu^+ \Pi_{WW}^{\mu\nu}(q^2) W^-_\nu\,,
\label{eq:vac_pol}
\end{equation}
where the (transverse) correlators can be decomposed as 
\begin{equation}
\Pi_{ij}^{\mu\nu}(q^2)=\left(\frac{q^{\mu}q^{\nu}}{q^2}-g^{\mu\nu}\right)\Pi_{ij}(q^2)\,.
\end{equation}
One defines the $\hat S \equiv \alpha S / 4 \sin^2 \theta_W$ parameter as the oblique contribution 
\begin{equation}
\hat S=\left(\frac{g}{g^{\prime}}\right)\Pi_{30}^{\prime}(0)
\end{equation}
and the $\hat T \equiv \alpha T$ parameter as  
\begin{equation}
\hat T = \frac{\Pi_{33}(0)-\Pi_{WW}(0)}{m_W^2}\,,
\end{equation}
{\it{i.e.}}, proportional to $\rho-1$ and therefore a measure of custodial symmetry breaking.

In the Standard Model, $S$ and $T$ receive contributions only at one-loop level (shown in Fig.~\ref{fig:1}). Their expressions are, in terms of the effective cut-off scale $\Lambda$,
\begin{subequations}
\begin{eqnarray}
\label{eq:SIR}\Delta \hat S_{IR} &=& \frac{g^2}{192 \pi^2} \log \frac{\Lambda^2}{m_W^2}\,, \\ 
\label{eq:TIR}\Delta \hat T_{IR} &=& - \frac{3\alpha}{16 \pi c_W^2} \log \frac{\Lambda^2}{m_W^2}\,. 
\end{eqnarray}
\end{subequations}
Residual finite terms are fixed by matching $\Delta \hat S$ and $\Delta \hat T$ to the SM reference point chosen in Ref.\cite{:2005ema} (with $m_t = 175 $GeV and $m_H = 150 $GeV), correcting for the most recent top mass value $m_t = 172$GeV and identifying the leading $m_H$ logarithmic dependence with that of $\Lambda$.

The starting point of the EWPO analysis in Higgsless models is the correlation between $S$ and $T$. Fixing the cutoff $\Lambda$ at the TeV scale, the main challenge of Higgsless models is to compensate the large negative infrared effect in $T$. However, when one includes spin-1 resonances, they also induce a tree level contribution to $S$, which is often assumed to be positive but is generically rather large in magnitude. 

In~\cite{Barbieri:2008cc}, an analysis was performed using an effective low energy description of resonance dynamics including the full $T$ at one loop. It was found that quadratically ultraviolet sensitive one-loop contributions of moderately light resonances can compensate the logarithmically divergent infrared effect in Eq.~(\ref{eq:TIR}) while giving a moderately large $S$, still within experimental bounds. 

However, the analysis of Ref.~\cite{Barbieri:2008cc} only considers resonance contributions to $S$ at tree level. Since there is no reason to expect small one-loop contributions to $S$, conclusions on the resonance spectrum and cut-off interplay in EWPT based on a correlated treatment of $S$ and $T$ may be significantly modified once the one-loop corrections to $S$ are included. Especially so because loop corrections presumably induce UV sensitivity to $S$ as well, skewing such correlations in a substantial way. Thus, a consistent correlated treatment of $S$ and $T$ should include quantum corrections on both of them. In the present study we extend the EWPO analysis of Ref.~\cite{Barbieri:2008cc} by including the full one-loop resonance contribution to $S$. 

This paper is organized as follows: in Section~\ref{sec:compo} we present the Lagrangian for the Higgsless model. The one-loop computation of $S$ is addressed in Section~\ref{sec:S}. Section~\ref{sec:analysis} is devoted to studying the modification of EWPT constraints on the parameter space of low lying resonances in Higgsless models and the interplay between UV sensitivity and the low lying resonance spectrum. Finally, some implications for direct resonance searches at colliders are discussed.

\begin{figure}[t]
\begin{center}
\includegraphics[width=5.cm]{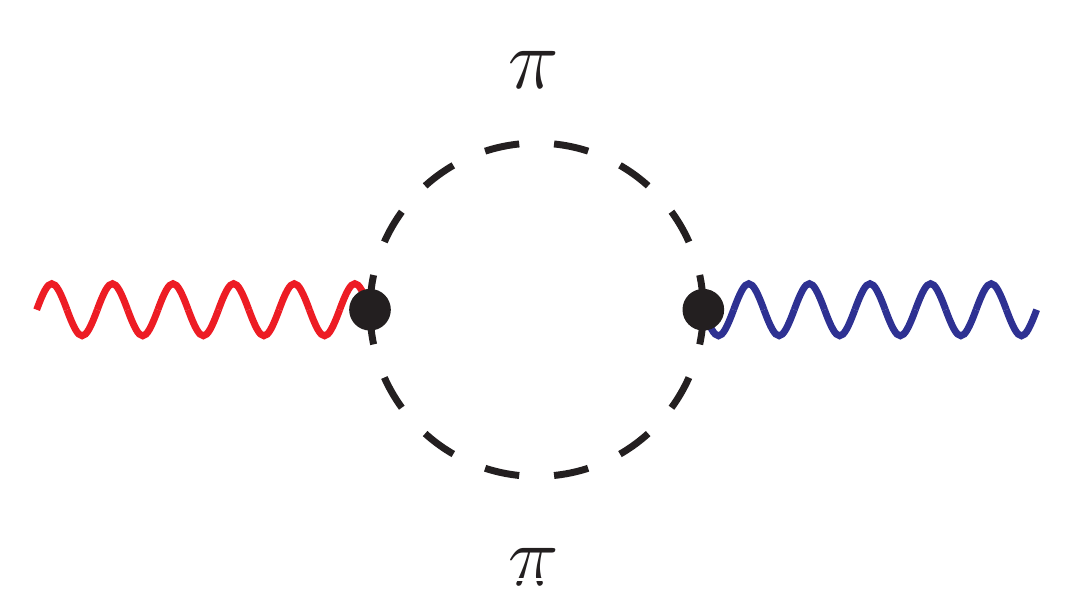}\hskip 0.2cm \includegraphics[width=10.cm]{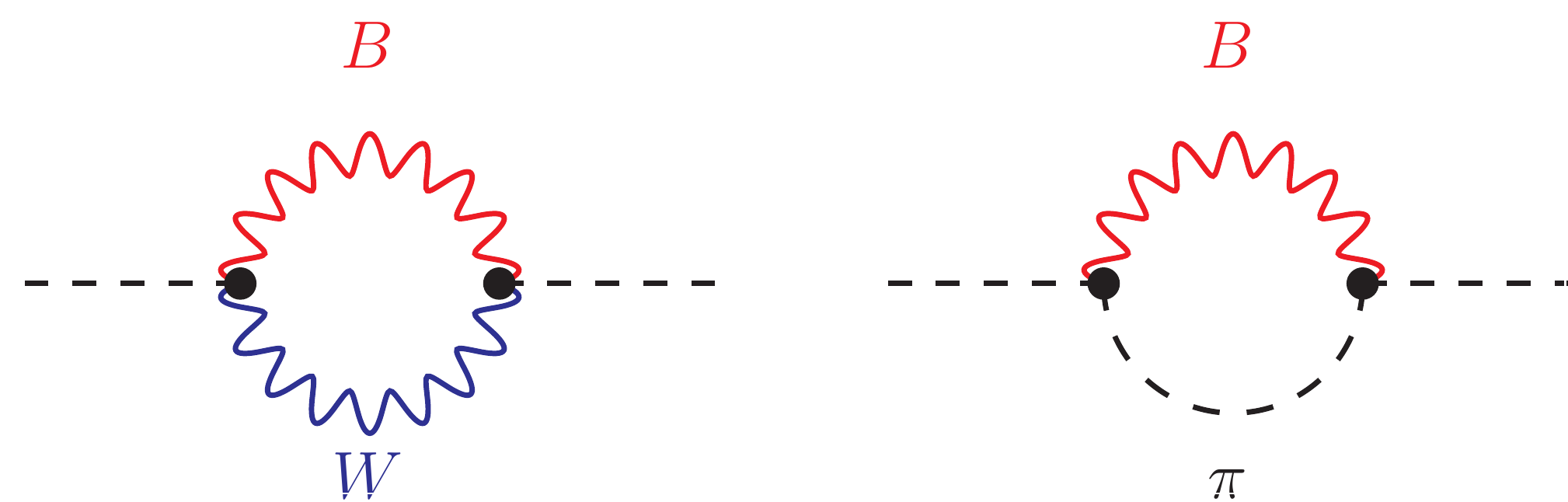}
\end{center}
\caption{\label{fig:2} {\it{{\footnotesize{Diagrams contributing to the infra-red contribution to $S$ (first diagram on the left hand side) and $T$.}}}}}
\end{figure}
%%%%%%%%%%%%%%%%%%%%%%%%%%%%%%%%%%%%%%%%%%%%%%%%%%%%
\section{The Lagrangian}\label{sec:compo}
%%%%%%%%%%%%%%%%%%%%%%%%%%%%%%%%%%%%%%%%%%%%%%%%%%%%
Our starting point will be a theory symmetric under $G=SU(2)_L\times SU(2)_R$, spontaneously broken down to the diagonal group $H=SU(2)_{L+R}$. The resulting Goldstone boson fields are elements of the coset $G/H$ and transform as 
\begin{equation}
u\rightarrow g_{R}uh^{\dagger}=hug_{L}^{\dagger}~,\qquad U\rightarrow g_{R}Ug_{L}^{\dagger}~,
\end{equation}
where $h=h(u,g_{L},g_{R})$ is an element of  $SU(2)_{L+R}$. It is customary to choose the following non-linear realization
\begin{align}
&  U=u^2=e^{i2\hat{\pi}/v},\qquad\hat{\pi}=T^{a}\pi^{a}=\frac{1}{\sqrt{2}}\left[
\begin{array}
[c]{cc}%
\frac{\pi^{0}}{\sqrt{2}} & \pi^{+}\nonumber\\
\pi^{-} & -\frac{\pi^{0}}{\sqrt{2}}%
\end{array}
\right]  ~,\qquad T^{a}=\frac{1}{2}\sigma^{a}\,.
\end{align}
The lowest-order Lagrangian one can build is the usual chiral Lagrangian
\begin{equation}
\mathcal{L}_{\chi}^{(2)}(U)=\frac{v^{2}}{4}\langle D_{\mu}U(D^{\mu}%
U)^{\dagger}\rangle\, , \label{eq:L2}%
\end{equation}                
where the covariant derivative contains the $W$ and $B$ fields:
\begin{align}
&  D_{\mu}U=\partial_{\mu}U-i\hat{B}_{\mu}U+iU\hat{W}_{\mu}~,\qquad\hat
{W}_{\mu}=gT^{a}W_{\mu}^{a}~,\qquad\hat{B}_{\mu}=g^{\prime}T^{3}B_{\mu}~,
\label{def}%
\end{align}
and $\langle\rangle$ denotes the trace of the $2\times2$ matrices. Since we want to avoid tree level contributions to $T$, explicit custodial symmetry breaking terms in Eq.~(\ref{eq:L2}) are omitted.

It is worth emphasizing that Eq.~(\ref{eq:L2}) is universal, {\it{i.e.}}, it does not depend on additional (heavy) degrees of freedom. Therefore, it provides by itself a mechanism for EWSB which is perfectly consistent: the Goldstone bosons become the longitudinal components of the $W^{+}, W^{-}, Z$ gauge bosons, thereby providing (tree level) masses proportional to $v$ with the exact same expressions as in the SM. The difference with the Higgs mechanism arises at the one-loop level -- at the NLO in the chiral expansion. We have already emphasized that Higgsless models are in general non-renormalizable: Eq.~(\ref{eq:L2}) is only the first term in a chiral expansion in powers of momenta while NLO terms in the chiral expansion capture ultraviolet effects (typically heavy resonances) in the most general way. 

In Higgsless models one often assumes that the new degrees of freedom are spin-1 resonances. We will consider the presence of two vector states with opposite parity, $A^{\mu\nu}[1^{++}]$ and $V^{\mu\nu}[1^{--}]$, both belonging to the adjoint representation of
$SU(2)_{L+R}$:
\begin{equation}
R^{\mu\nu}\rightarrow hR^{\mu\nu}h^{\dagger}~,\qquad R^{\mu\nu}=A^{\mu\nu
},\ V^{\mu\nu}~. \label{eq:Rtr}%
\end{equation}
For phenomenological purposes we will assume $v< m_{V,A}< \Lambda\simeq 4\pi v$. In other words, we consider two relatively light resonances and make the natural cutoff $\Lambda$ of the theory (signaling the presence of new heavy degrees of freedom) to coincide with the loop expansion parameter $\Lambda_{\chi}=4\pi v$.

Following Ref.~\cite{Ecker:1988te}, we describe the heavy spin-1 states in the antisymmetric tensor representation. Unlike in the vector representation, this way one avoids an unwanted mixing between the axial resonance and the pion. The kinetic term in the Lagrangian has the form:
\begin{equation}
\mathcal{L}_{\mathrm{kin}}(R^{\mu\nu})=-\frac{1}{2}\langle\nabla_{\mu}%
R^{\mu\nu}\nabla^{\sigma}R_{\sigma\nu}\rangle+\frac{1}{4}m_{R}^{2}\langle
R^{\mu\nu}R_{\mu\nu}\rangle~,
\end{equation}
with the covariant derivative defined as
\begin{equation}
\nabla_{\mu}R=\partial_{\mu}R+[\Gamma_{\mu},R],\qquad\Gamma_{\mu}=\frac{1}%
{2}\left[  u^{\dagger}(\partial_{\mu}-i\hat{B}_{\mu})u+u(\partial_{\mu}%
-i\hat{W}_{\mu})u^{\dagger}\right]  ,\quad\Gamma_{\mu}^{\dagger}=-\Gamma_{\mu
}\,.
\end{equation}
In order to couple the heavy states to the Goldstone bosons it is convenient to define $u_{\mu}=iu^{\dagger}D_{\mu}Uu^{\dagger}=u_{\mu}^{\dagger}$, such that $u_{\mu}\rightarrow hu_{\mu}h^{\dagger}$.
  
The leading order, $\mathcal{O}(p^{2})$, couplings of the heavy fields to
Goldstone bosons and SM gauge fields are then parametrised in terms of 
3 effective operators, defined by 
\bea
\mathcal{L}_{1V}^{(2)} &=& \frac{i}{2\sqrt{2}}G_{V}\langle V^{\mu\nu}[u_{\mu
},u_{\nu}]\rangle +\frac{1}{2\sqrt{2}}F_{V}\langle V^{\mu\nu}(u\hat{W}^{\mu\nu
}u^{\dagger}+u^{\dagger}\hat{B}^{\mu\nu}u)\rangle \no \\
&+&\frac{1}{2\sqrt{2}}F_{A}\langle A^{\mu\nu}(u\hat{W}^{\mu\nu}u^{\dagger}-u^{\dagger}\hat{B}%
^{\mu\nu}u)\rangle\,. \label{eq:LV}%
\eea
 
The effective couplings $F_{V,A}$ and $G_V$ have dimensions of mass and, by naive dimensional analysis, are expected to be of $\cO(v)$. More quantitative constraints can be found by requiring the resonance contribution to smoothen the ultraviolet divergences generated by the leading order term, Eq.~(\ref{eq:L2}), in a set of chosen correlators, in the spirit of~\cite{Ecker:1989yg}. For instance, perturbative unitarity in $W_L W_L \to W_L W_L$ scattering up to the natural cut-off scale of the model $\Lambda \simeq 4\pi v \approx 3$~TeV imposes the constraint $G_V \simeq v / \sqrt 3(1\pm0.2)$, mostly independent of the vector mass below $m_V< 1$~TeV~\cite{Barbieri:2008cc}. Furthermore, the pion vector form factor, in combination with positivity of $F_V, G_V$ requires $F_V \lesssim v^2/G_V$. Combined with the unitarity constraint on $G_V$ it gives $F_V \lesssim  \sqrt 3 v$~\cite{Barbieri:2008cc}. Additionally, imposing the right ultraviolet behavior of $\Pi_{30}$ (and neglecting heavier resonance contributions) suggests that $m_A>m_V$ and also $F^2_V/m^2_V > F_A^2 / m_A^2$, which are also known as the Weinberg sum rules.

Another constraint which is typically discussed in Higgsless models is the ``hidden-gauge relation'' $F_{V}=2G_V$. This relation holds in 3- and 4-site Higgsless models~\cite{Barbieri:2008cc}, and more generally in all five-dimensional deconstructed models as long as non-renormalizable terms in the bulk are neglected~\cite{Hirn:2007we}. In more general Higgsless frameworks, however, it is typically not satisfied. In the following sections we will check whether it is consistent with EWPO.

Finally, let us stress that the interaction Lagrangian of Eq.~(\ref{eq:LV}) is only the most general $\cO(p^2)$ chiral Lagrangian {\emph{linear}} in resonance fields. Note that since there is no power counting that supresses $V_{\mu\nu}$ and $A_{\mu\nu}$, operators with an arbitrary number of resonance fields have the same chiral counting and appear already at ${\cal{O}}(p^2)$. The predictability of the approach therefore rests on the assumption that those extra operators do not contribute significantly. We will return to this issue later on. 

%%%%%%%%%%%%%%%%%%%%%%%%%%%%%%%%%%%%%%%%%%%%%%%%%%%%
\section{S and T at one loop}\label{sec:S}
%%%%%%%%%%%%%%%%%%%%%%%%%%%%%%%%%%%%%%%%%%%%%%%%%%%%

The tree level exchange of the heavy vectors (Fig.~\ref{fig:3}) leads to well known contribution to $S$ (and also $W, Y$~\cite{Barbieri:2004qk}) whereas it leaves $T$ untouched because of the protecting $SU(2)_{L+R}$ symmetry:
\begin{figure}[t]
\begin{center}
\includegraphics[width=5.cm]{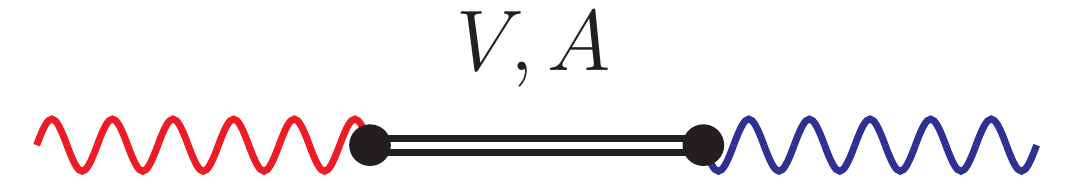}
\end{center}
\caption{\label{fig:3} {\it{{\footnotesize{Tree level resonance contribution to $S$.}}}}}
\end{figure}
\begin{subequations}
\begin{eqnarray}
\Delta \hat S_{Tree} &=& \frac{g^2}{4} \left( \frac{F_V^2}{m_V^2} - \frac{F_A^2}{m_A^2} \right) \,, \\ 
\Delta \hat T_{Tree} &=& 0\,. 
\end{eqnarray}
\end{subequations}
The contribution to $S$ can in principle be positive or negative, although as discussed in the previous section, positive values are preferred to comply with the right ultraviolet behavior of $\Pi_{30}$. This leads to the well-known problem of Higgsless models with EWPT, since tree level resonance contributions actually worsen the predictions due to the pure infrared effects of Eqs.~(\ref{eq:SIR}, \ref{eq:TIR}). 

At one loop one has to distinguish between the long distance and the short distance contributions. The former are depicted in Fig.~\ref{fig:2} and, since they come entirely from Eq.~(\ref{eq:L2}), they coincide with the SM result (with the Higgs mass instead of $\Lambda$) and are given entirely by Eqs.~(\ref{eq:SIR}, \ref{eq:TIR}). 

\begin{figure}[t]
\begin{center}
\includegraphics[width=9.cm]{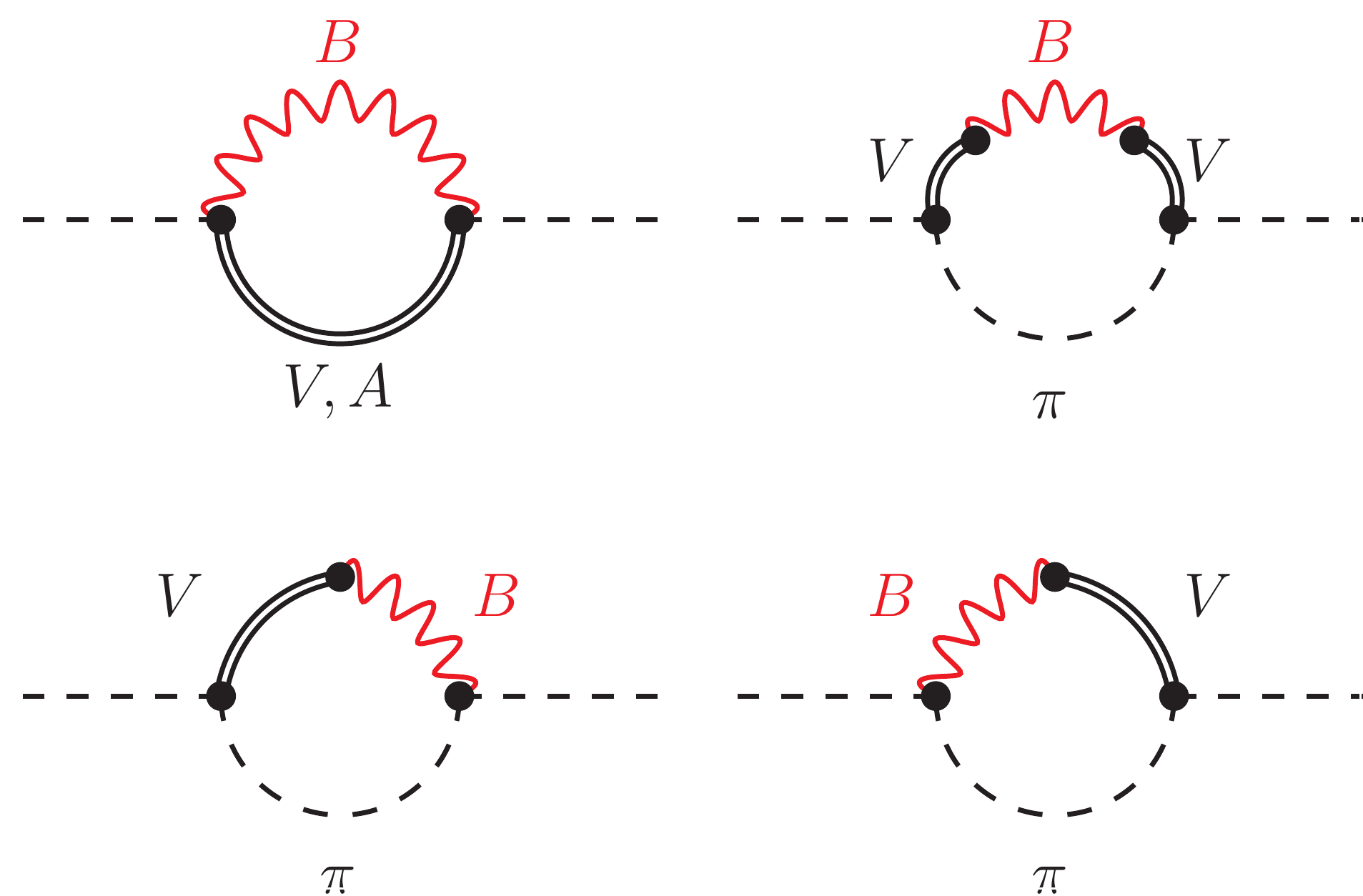}
\end{center}
\caption{\label{fig:4} {\it{{\footnotesize{Short distance contributions to $T$. The first diagram contains quadratic divergences, while the rest only diverge logarithmically.}}}}}
\end{figure}

One-loop short distance contributions to ${T}$ involve the diagrams with resonance exchange shown in Fig.~\ref{fig:4}. The calculation of loop contributions to $T$ can be significantly simplified by noticing that, up to corrections of relative order $m^2_W / m_R^2$
\begin{equation}
\hat T = \frac{Z^{(+)}}{Z^{(0)}}-1\,,
\end{equation}
where $Z^{(+)}$, $Z^{(0)}$ are the wave-function renormalization constants of the charged and neutral Goldstone bosons for the SM gauge bosons (in the Landau gauge)~\cite{Barbieri:2007gi}.\footnote{Contributions to $U$ can be computed analogously. However, since resonance contributions to $U$ are suppressed by $v^2/m_R^2$ with respect to $T$, we do not consider them in our numerical analysis. The same applies to the contributions to $W, X, Y$ with respect to $S$.}

\begin{figure}[t]
\begin{center}
\includegraphics[width=10.cm]{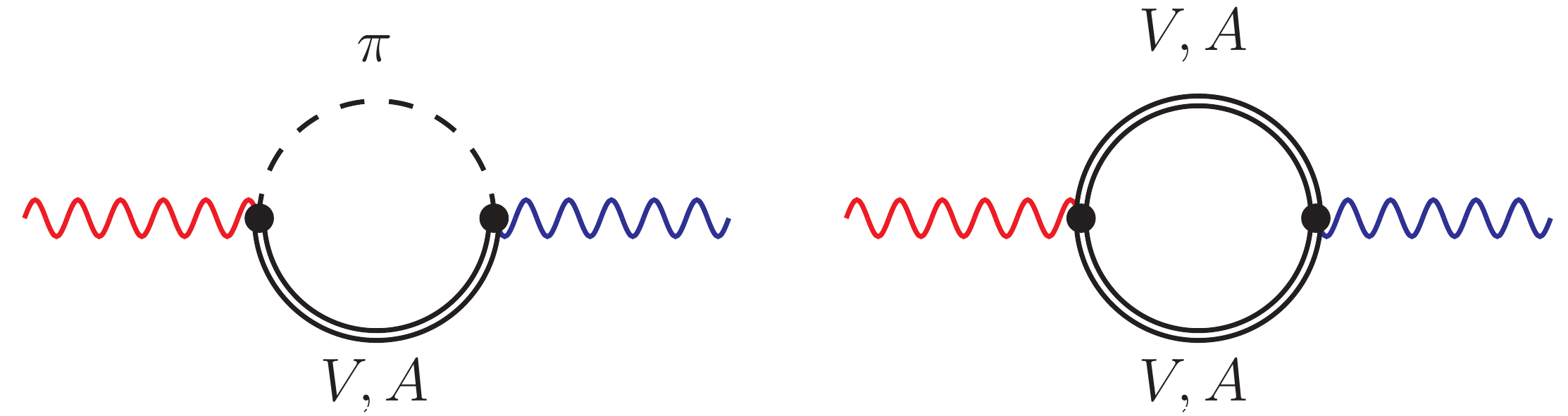}
\end{center}
\caption{\label{fig:5} {\it{{\footnotesize{Diagrams yielding resonance contributions to $S$ at one-loop order.}}}}}
\end{figure}

\begin{figure}[h!]
\begin{center}
\includegraphics[width=12.cm]{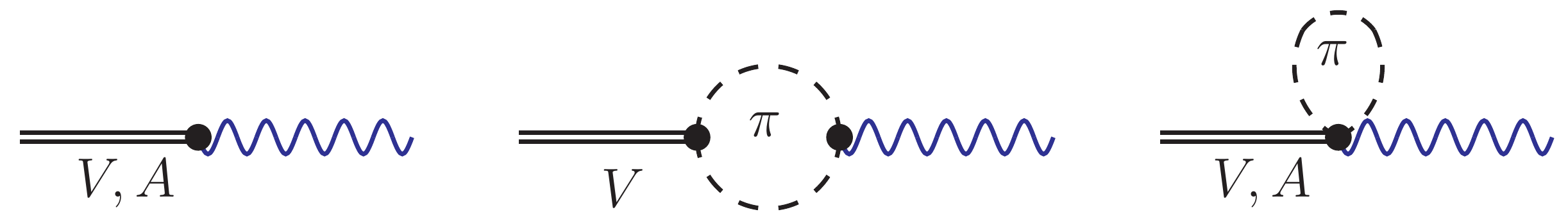}
\end{center}
\caption{\label{fig:6} {\it{{\footnotesize{Diagrams contributing to the renormalization of $F_V$ and $F_A$.}}}}}
\end{figure}

$S$ receives one-loop short distance contributions from the diagrams depicted in Fig.~\ref{fig:5}. In addition, renormalization of the resonance masses and couplings entering the tree level contribution to $S$ needs to be taken into account. It is easy to verify that the masses do not get renormalized at this order while the corrections to the $F_{V,A}$ couplings, depicted in Fig.~\ref{fig:6}, amount to the shifts
\begin{eqnarray}
F_V&\to& F_V\left(1+\frac{F_V-G_V}{F_V}\frac{\Lambda^2}{(4\pi v)^2}\right)\,,\nonumber\\
F_A&\to& F_A\left(1+\frac{\Lambda^2}{(4\pi v)^2}\right)\,.\label{eq:FVA}
\end{eqnarray}
Notice that the corrections are purely quadratic in $\Lambda$ and that no logarithmic terms are induced. The presence of quadratically (ultraviolet) divergent pieces in the one-loop contributions to $S$ and $T$ will play a paramount role in our results and will be discussed in short. For $T$ they come only from the top left hand side diagram in Fig.~\ref{fig:4}, while for $S$ in addition to the coupling corrections in (\ref{eq:FVA}), they are also contained in both diagrams in Fig.~\ref{fig:5}.
 

The final expressions for the short distance contributions to both $S$ and $T$ read
\begin{eqnarray}\label{UV}
\Delta \hat S_{UV} &=& \frac{g^2}{4} \frac{\Lambda^2}{(4\pi v)^2} \left[\frac{2G_V(F_V-2G_V)}{m_V^2}+ \left( \frac{F_V}{m_V} \right)^2- \left( \frac{F_A}{m_A}\right)^2 - \frac{2}{3}\left(\frac{v^2}{m_V^2} + \frac{v^2}{m_A^2}\right) \right]\nonumber\\
&+& \frac{g^2}{4}\frac{1}{(4\pi v)^2}\left[\left(\frac{10}{3}G_V^2-F_V^2-v^2\right)\log\frac{\Lambda^2}{m_V^2}+(F_A^2-v^2)\log\frac{\Lambda^2}{m_A^2}\right]\,, \\ 
\Delta \hat T_{UV} &=&  \frac{3\pi \alpha}{4c_W^2} \frac{\Lambda^2}{(4\pi v)^2}  \left[ \left( \frac{F_V-2G_V}{m_V}\right)^2  + \left( \frac{F_A}{m_A}\right)^2 \right] \,\nonumber\\
\label{eq:17}&+&  \frac{3\pi\alpha}{2 c_W^2(4\pi v)^2}\left[ \left(2G_V^2+4F_VG_V-F_V^2\frac{2G_V^2+v^2}{v^2}\right) \log \frac{\Lambda^2}{m_V^2} - F_A^2 \log \frac{\Lambda^2}{m_A^2}  \right] \,,\nonumber\\  
\end{eqnarray}
where we have included the quadratic and the leading logarithmic contributions, and the expression for $S$ already contains the shift due to $F_{V,A}$ renormalization. We checked that our results for the logarithmic contribution to $\hat S$ agree with those found for $L_{10}$ in ChPT with explicit resonance fields~\cite{Cata:2001nz}.\footnote{Note that one has to correct for the fact that the result in Ref.~\cite{Cata:2001nz} is done in $SU(3)_L\times SU(3)_R$ ChPT.} Regarding $\hat T$, we have checked that in the hidden gauge limit our result agrees with~\cite{Barbieri:2008cc}.\footnote{We are thankful to Axel Orgogozo and Slava Rychkov for pointing out an error in Eq.~\eqref{eq:17} in the first version of the paper~\cite{Orgogozo:2011kq}. }
%note that the second term in square brackets is missing in Eq.~(4.6) of Ref.~\cite{Barbieri:2008cc}. It originates from the logarithmically divergent part of the upper right hand side diagram in Fig.~\ref{fig:4} and it would modify their first term in curly brackets to $\log(m_V/m_W) [1+(G_V/\sqrt 3 v)^2]$. Numerically this gives a negligible $\sim 10 \%$ correction to the infra-red logarithmic coefficient for the unitarity preferred values of $G_V$.

The first thing to note is that the quadratic contribution to $\hat T$ is strictly positive, while the new contribution to $\hat S$ can in principle have either sign. In addition, the quadratic pieces are parametrically large, of the same order of magnitude as the tree level contributions. In principle it would then be desirable to find a way to cancel (or at least ameliorate) their size. 
\begin{figure}[t]
\begin{center}
\includegraphics[width=10cm]{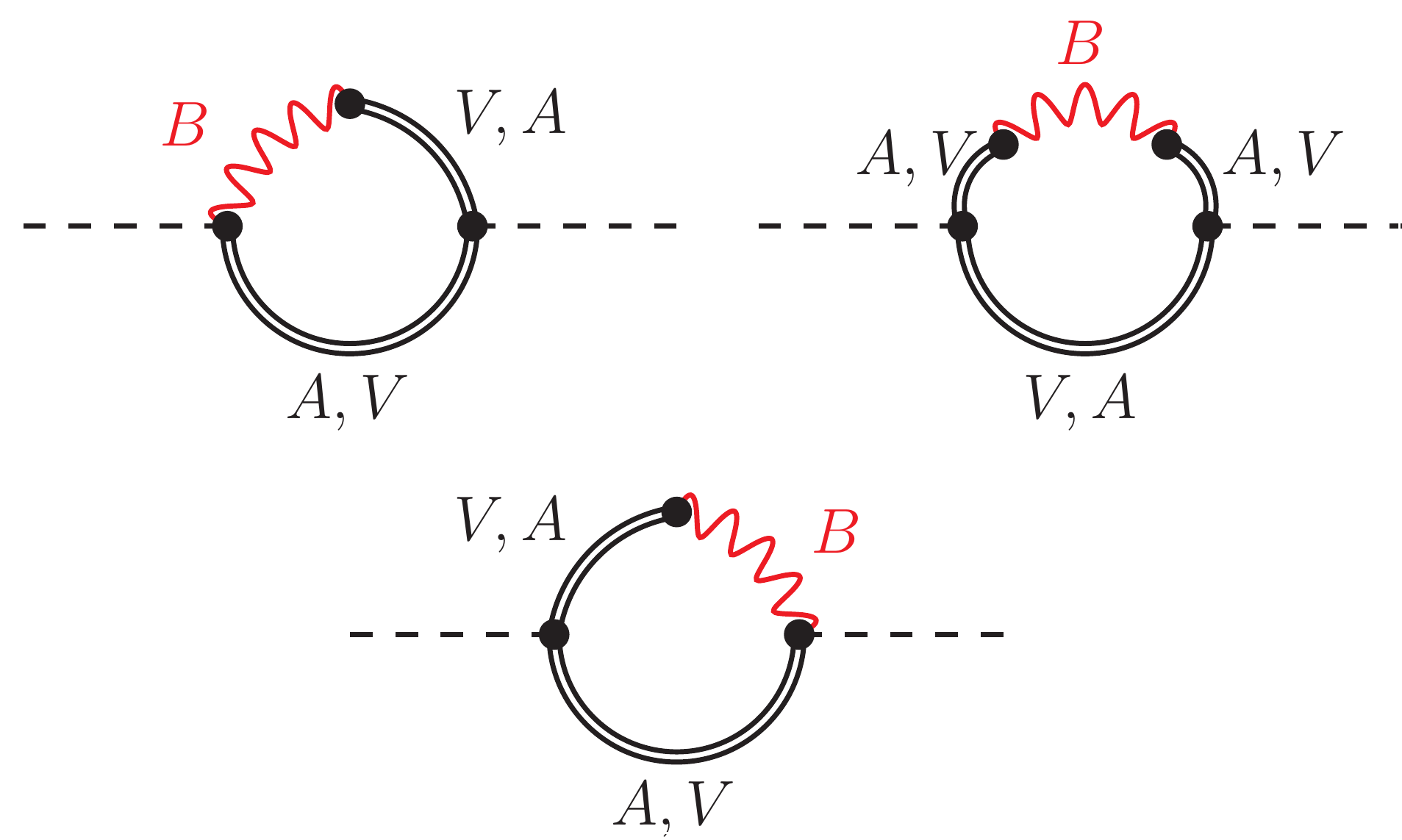}
\end{center}
\caption{\label{fig:9} {\it{{\footnotesize{Contributions cutting off quadratic divergences to $\hat T$.}}}}}
\end{figure}

One possible mechanism is to impose the ``hidden gauge" limit relations, $F_V\to 2 G_V$, $F_A\to 0$ and  $m_A \to \infty$, which as we noted earlier arise naturally in a wide class of Higgsless models. The quadratic dependence in $\hat T$ then identically cancels and we are left only with the logarithmic piece, whose sign is in principle not fixed.  While this feature of the hidden-gauge limit for $\hat T$ is very appealing, it was already noticed in~\cite{Barbieri:2008cc} that the hidden-gauge relation conflicts with EWPO. Our calculation here shows that it also fails to protect $\hat S$, which still scales quadratically due to the resonance loop in the right hand side diagram of Fig.~\ref{fig:5}. In particular, this implies that for the unitarity preferred values of $G_V$ and the natural cut-off scale, the one-loop contribution to $\hat S$ is half the tree level value. From Eq. (\ref{eq:FVA})  it also follows that the hidden gauge limit is not necessarily respected by the renormalization of model parameters.

The alternative is to take the quadratic divergences as a physical effect. In Ref.~\cite{Barbieri:2008cc} it was shown that the introduction of the following new operators bilinear in the resonance fields
\bea
\langle A^{\mu\nu}[\nabla_{\rho}V^{\rho\nu},u_{\mu}]\rangle
\,;\quad \langle V^{\mu\nu}[\nabla_{\rho}A^{\rho\nu},u_{\mu}]\rangle~,\label{eq:L2V}
\eea
can cancel the quadratic divergences and tame the logarithmic ones. This contribution is shown in Fig.~\ref{fig:9}. Therefore, if the renormalization scale happens to be due to masses lying close to $\Lambda$, then the expression for $T$ in Eqs.~(\ref{UV}) should be a reasonable estimate.  

\begin{figure}[t]
\begin{center}
\includegraphics[width=17cm]{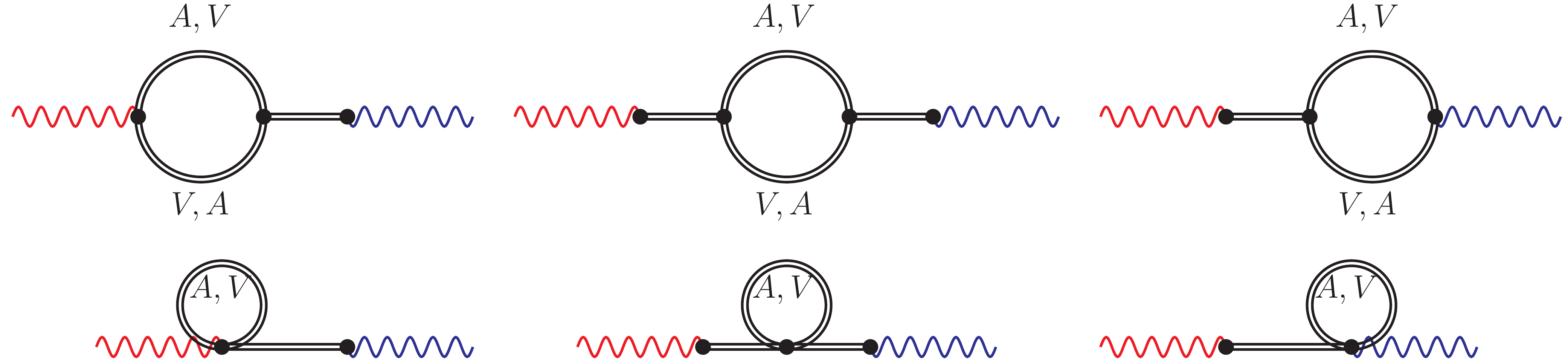}
\end{center}
\caption{\label{fig:10} {\it{{\footnotesize{Some contributions expected to regulate the quadratic divergences in $\hat S$.}}}}}
\end{figure}

The same mechanism in principle could apply for $\hat S$, but the situation becomes more involved. First, we have explicitly checked that the contribution of the operators of Eq.~(\ref{eq:L2V}) not only does not cancel the quadratic divergences in $\hat S$ but introduces quartic ones. More operators are therefore needed. For instance, we expect operators with 3 and 4 resonance fields of the form 
\begin{equation}
\langle[\nabla^{\mu}V_{\mu\nu},\nabla^{\lambda}V_{\lambda\rho}]V^{\nu\rho}\rangle~;\quad \langle \nabla^{\mu}V_{\mu\nu}(V^{\eta\nu}\nabla_{\lambda}V^{\lambda\rho}V_{\eta\rho}+V_{\eta\rho}\nabla_{\lambda}V^{\lambda\rho}V^{\eta\nu})\rangle~,\label{eq:L2extra}
\end{equation}
to play a significant role. In particular, their contribution to the diagrams in Fig.~\ref{fig:10} is expected to compensate the divergences coming from the right-hand side diagram in Fig.~\ref{fig:5}. The list of operators in Eq.~(\ref{eq:L2extra}) is however far from being exhaustive. For instance, one should also include terms mixing axial and vector fields to compensate for the divergences generated by the operators in Eq.~(\ref{eq:L2V}). In general, such additional operators will also affect $\hat T$.   

While the full renormalization analysis is clearly beyond the scope of the present paper, we want to point out that it is in principle feasible. Even though there is no well-defined power counting for the resonance fields, if one assumes the presence of a single vector and axial resonance field, the one-loop renormalization of $\hat S$ and $\hat T$ and in general every EWPO can be accounted for with operators with up to 4 resonance fields. A reasonable strategy is then to assume that indeed there is a combination of operators able to regulate the quadratic (and quartic) divergences to both $\hat S$ and $\hat T$ and that the scale at which this happens is close to $\Lambda$. However, there is a potential problem with such an assumption. If the renormalization scale for all diagrams is around $\Lambda$ the corrections to $F_V$ and $F_A$ are not necessarily well behaved perturbatively. Consequently, the values for $F_V$ and $F_A$ coming from the fit to EWPO may be unreliable. An alternative approach would be to require that $F_{V,A}$ renormalization in Fig.~\ref{fig:6} is governed by the resonance scale, while the diagrams of Figs.~\ref{fig:4} and~\ref{fig:5} still get renormalized at the cutoff scale (in order to be consistent with EWPO). While such a scenario cannot be excluded, it would make the model unnaturally baroque. In the following section we will show that fortunately even for a single renormalization scale close to $\Lambda$, the correlations between $S$ and $T$ tend to stabilize the values of $F_V$ and $F_A$.   

%%%%%%%%%%%%%%%%%%%%%%%%%%%%%%%%%%%%%%%%%%%%%%%%%%%%
\section{Confronting EWPT and Unitarity Constraints}\label{sec:analysis}
%%%%%%%%%%%%%%%%%%%%%%%%%%%%%%%%%%%%%%%%%%%%%%%%%%%%

\begin{figure}[t]
\begin{center}
  \includegraphics[height=.4\textheight]{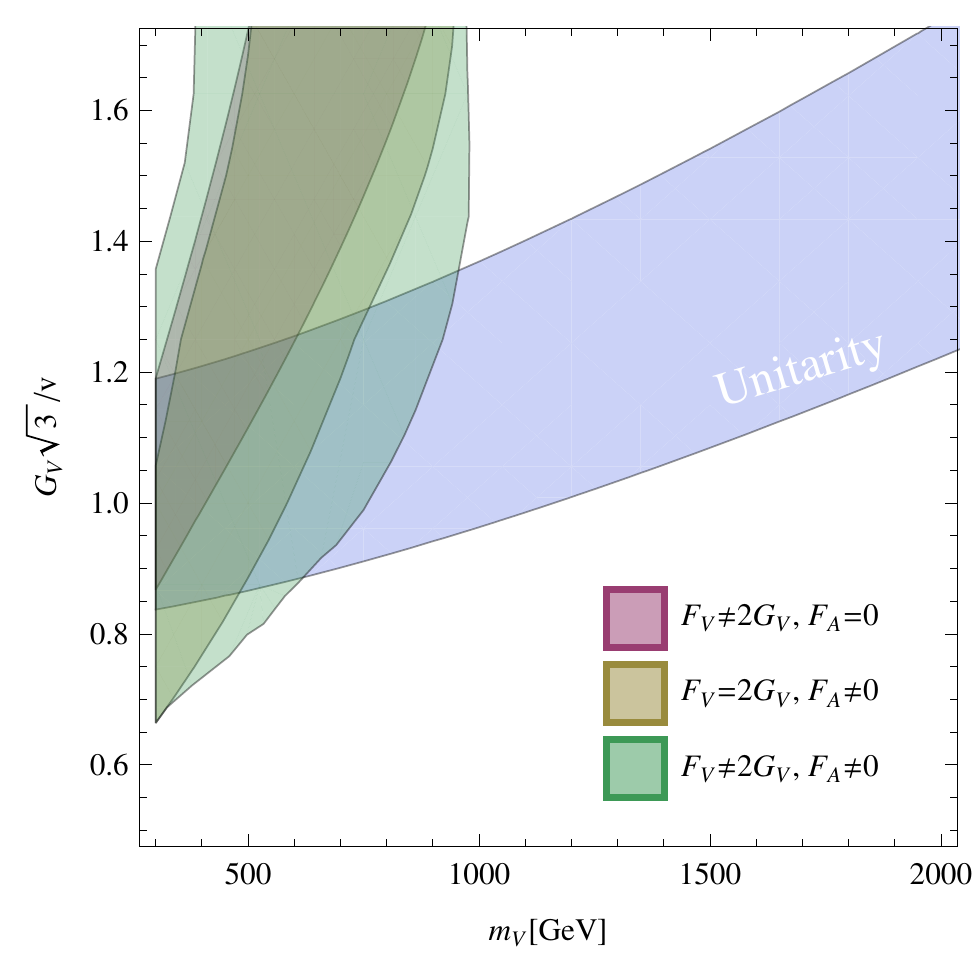}
\end{center}
  \caption{\label{fig:STU} {\it{{\footnotesize{Combination of unitarity (precise range taken as in Ref.\cite{Barbieri:2008cc}) and EWPT constraints (at 95\% C.L.) in the ($m_V$, $G_V$) plane. See text for details.}}}}}
\end{figure}

\begin{figure}
\begin{center}
  \includegraphics[height=.25\textheight]{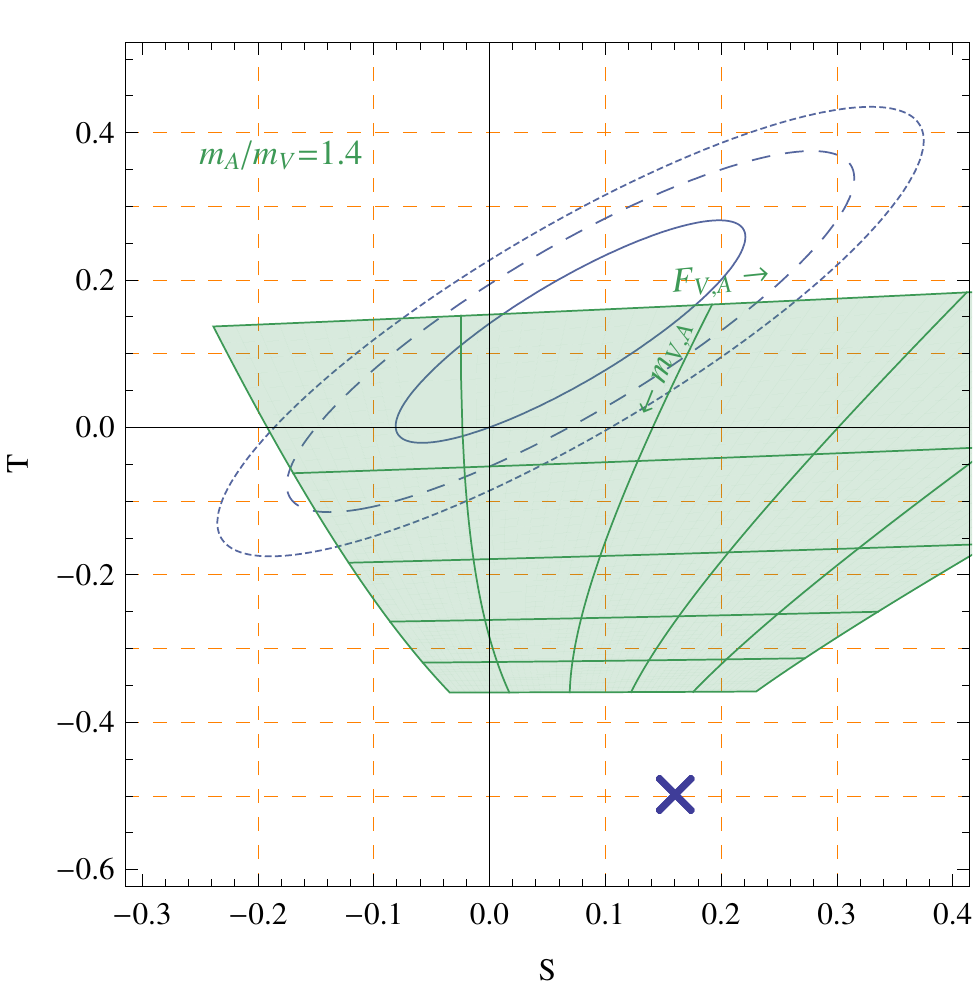}
  \includegraphics[height=.25\textheight]{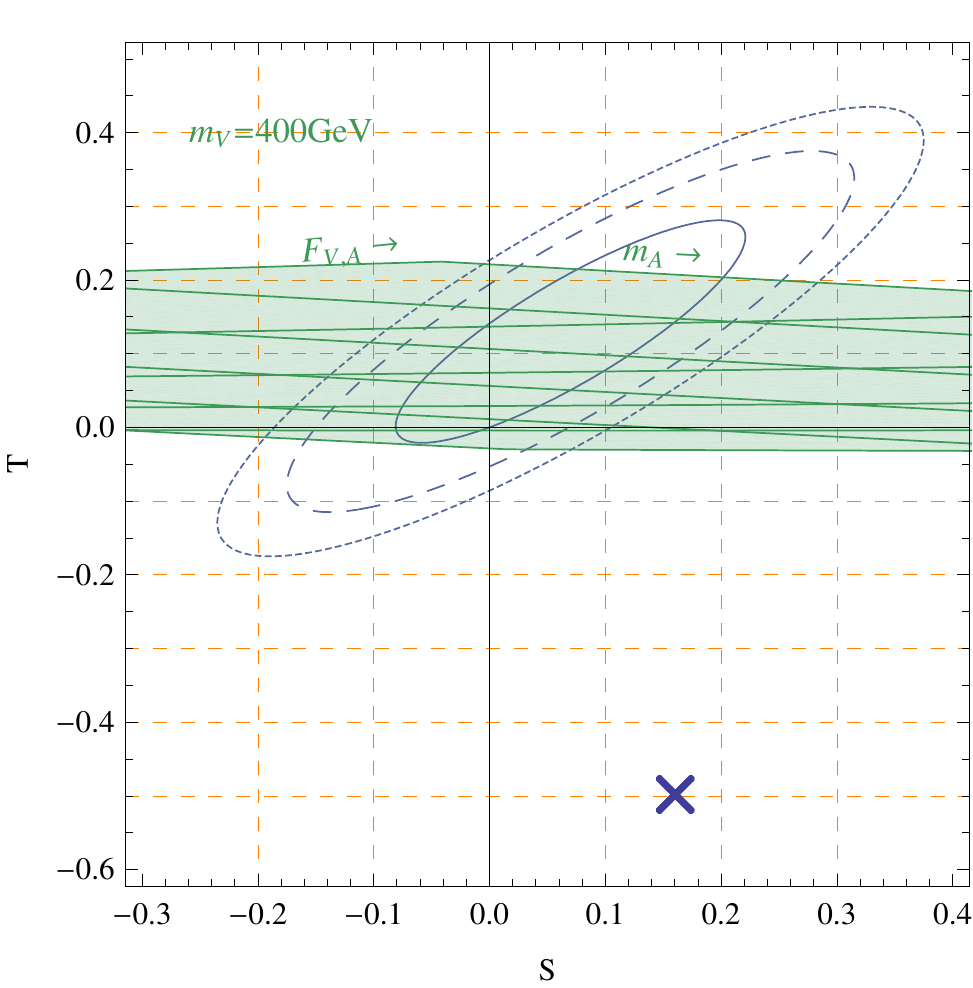}
  \includegraphics[height=.25\textheight]{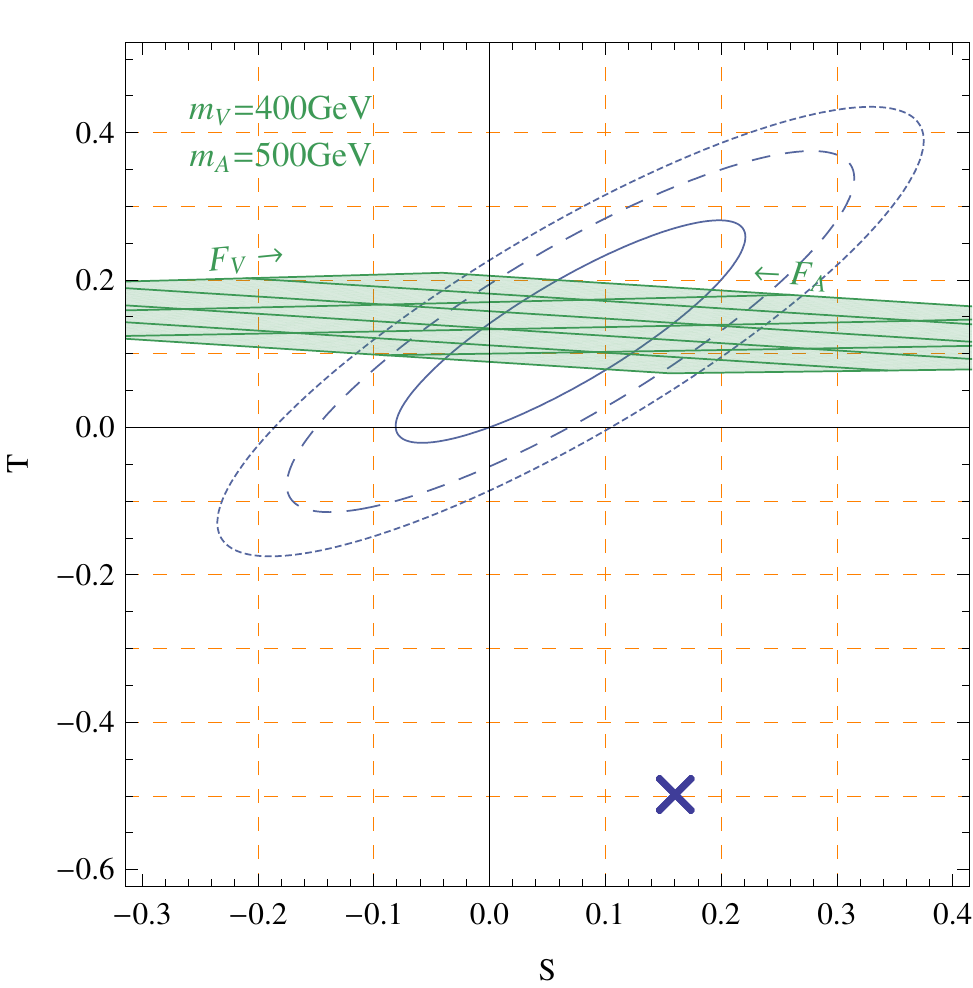}
\end{center}
  \caption{\label{fig:1c} {\it{{\footnotesize{Contributions to $\hat S$ and $\hat T$ in the scenario with a vector and an axial resonance below the cut-off of $\Lambda=3$~TeV. The exact unitarity constraint $G_V=v/\sqrt 3$ is imposed.  On the upper-left plot, $F_A=F_V$ and $m_A/m_V=1.4$ while $m_V$ and $F_V$ are varied in the ranges $500~\mathrm{GeV} \leq m_V \leq 1~\mathrm{TeV}$, $0.9\leq F_V/2G_V \leq 1$.  On the upper-right plot, again $F_A=F_V$ and vector mass is fixed at $m_V=400$~GeV. The remainig parameters are varied in the ranges $540~\mathrm{GeV} \leq m_A \leq 700~\mathrm{GeV}$, $0.8\leq F_V/2G_V \leq 0.95$\,. On the lower plot, the vector and axial masses are fixed at $m_V=400$~GeV and $m_A=500$~GeV, while the couplings are varied in the ranges $0.9\leq F_V/2G_V \leq 0.95$ and $0.8\leq F_A/2G_V \leq 0.9$.}}}}}
\end{figure}

We are now in the position to reconsider the interplay between the $W_L W_L$ scattering unitarity constraint and EWPT on the parameter space of Higgsless models at the one-loop level. We will study the impact of new contributions to $\hat S$ by reexamining the scenarios and conclusions reached in the original analysis of Ref.~\cite{Barbieri:2008cc}, where the resonance contributions to $\hat S$ were evaluated at the tree-level. The results are summarized in Fig.~\ref{fig:STU}, where we plot the allowed regions in the $G_V$ versus $m_V$ plane by marginalizing over the remaining free parameters.  
In order to assess the potential consistency of more specific classes of Higgsless models with EWPO we will consider the following scenarios (also marked in Fig.~\ref{fig:STU}):
\begin{itemize}
\item $F_V \neq 2 G_V$, $F_A=0$, $m_V<\Lambda$, $m_A\gsim \Lambda$, {\it{i.e.}}, a single vector resonance, not (necessarily) fulfilling the hidden gauge relation. In addition to the IR logarithmic contributions and the positive tree level contribution to $\hat S$, there are also quadratically UV sensitive one-loop contributions to both $\hat S$ and $\hat T$. They are strictly positive in $\hat T$ but can be negative in $\hat S$, thus allowing for potential cancellations between the tree level and one loop contribution. Interestingly, a consistent fit to both $\hat S$ and $\hat T$ can be achieved for light vector masses, $m_V\lesssim 500$~GeV, and reasonable values of the parameter $F_V\lesssim 2G_V$, as illustrated in Fig.~\ref{fig:1a}.
\begin{figure}[t]
\begin{center}
  \includegraphics[height=.33\textheight]{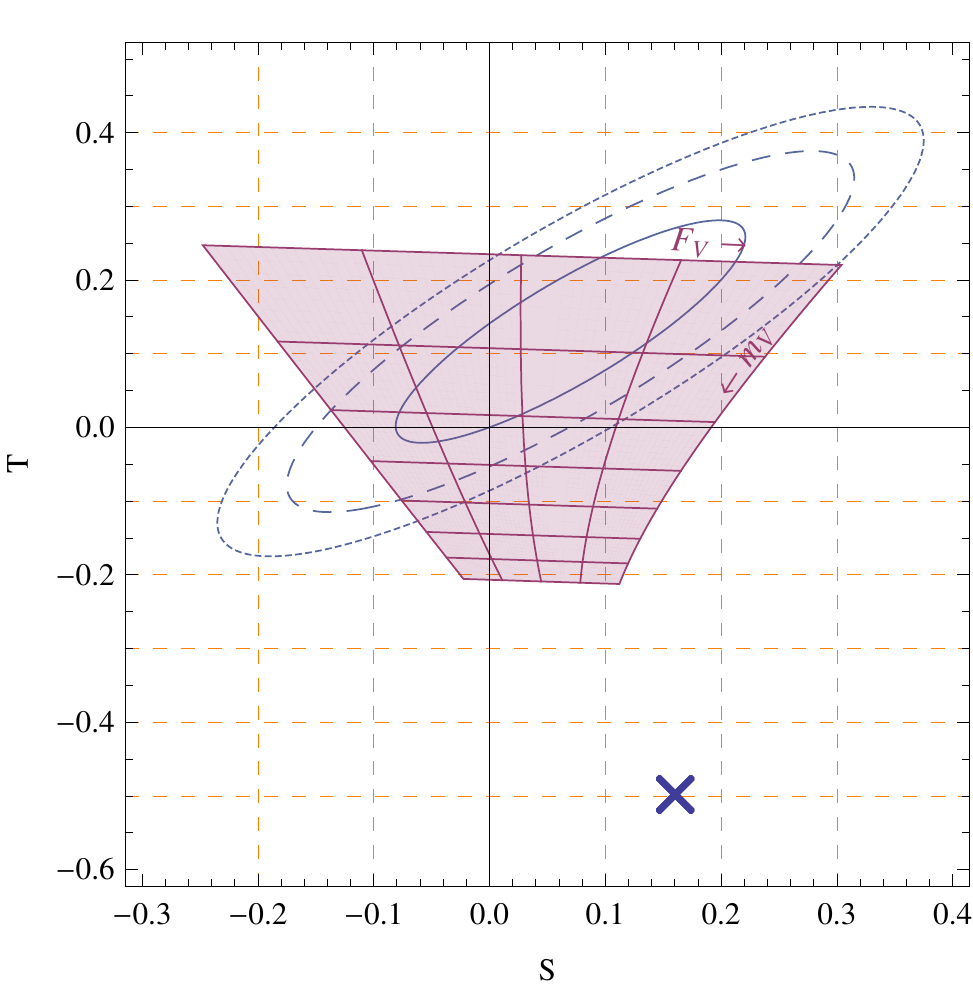}
\end{center}
  \caption{\label{fig:1a} {\it{{\footnotesize{Contributions to $\hat S$ and $\hat T$ in scenarios with a single vector resonance below the cut-off $\Lambda=3$~TeV. The exact unitarity constraint $G_V=v/\sqrt 3$ is imposed.  The parameters are varied over the ranges $250~\mathrm{GeV} \leq m_V \leq 500~\mathrm{TeV}$ and $0.64\leq F_V/2 G_V \leq 0.65$.}}}}}
\end{figure}
Comparing with Ref.~\cite{Barbieri:2008cc}, one observes that the main impact of the one-loop contribution to $\hat S$ is to lower the allowed vector resonance mass while raising the value of $F_V$. If one assumes that the value for $F_V$ is stable under quantum corrections, then this scenario is already in conflict with bounds from direct resonance searches at the Tevatron~\cite{Cata:2009iy}. This might well be the case: even though the operators we are considering induce a naive ${\cal{O}}(1)$ loop correction to $F_V$, their values extracted from the electroweak fit (with and without considering vertex corrections) only differ by roughly $15\%$. In other words, the EWPO fit seems to point to a stable region of parameters.
%%%%%%%%%%%%%%%%%%%%%%%%%%%%%%%%%%%%%%%%%%%%%%%%%%%%%
\item $F_V = 2 G_V$, $F_A\neq 0$, $(m_V,m_A)<\Lambda$, {\it{i.e.}}, a vector resonance fulfilling the hidden gauge relation plus an axial-vector resonance, both below the cut-off scale. An axial resonance is commonly considered in Higgsless models since its contribution reduces the size of $\Delta \hat S$ at tree level. At one loop it contributes with a quadratic UV sensitivity to both $\hat S$ and $\hat T$. We note that while the contribution to the latter is always positive, it can be either sign for the former, again opening the door for fine-tuned cancellations between $F_A$, $m_A$ and $m_V$. Figure~\ref{fig:1b} shows that a consistent fit to both $\hat S$ and $\hat T$ can only be achieved for relatively degenerate vector and axial masses below $700$~GeV. We note that $m_A$ can be made heavier at the expense of increasing $F_A$. In other words, even though the one loop expression introduces the new ratio $v/m_A$, apparently it does not break the correlation between $F_A$ and $m_A$ that one has at tree level.  
\begin{figure}
\begin{center}
  \includegraphics[height=.33\textheight]{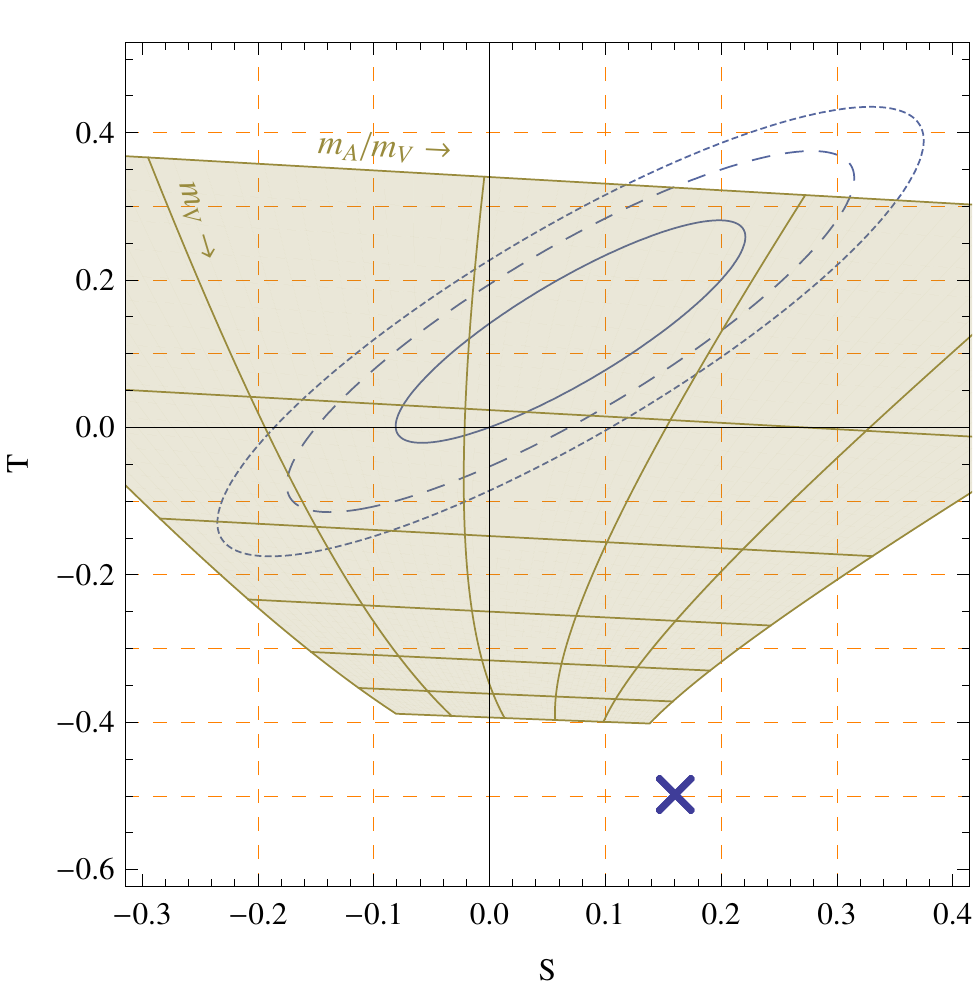}~~~
\end{center}
  \caption{\label{fig:1b} {\it{{\footnotesize{Contributions to $\hat S$ and $\hat T$ in the scenario with a vector resonance fulfilling $F_V=2G_V$ and an axial resonance below $\Lambda=3$~TeV. The exact unitarity constraint $G_V=v/\sqrt 3$ is imposed together with $F_A=F_V$. The masses are varied in the ranges $400~\mathrm{GeV} \leq m_V \leq 1~\mathrm{TeV}$, $1.25\leq m_A/m_V \leq 1.35$. 
}}}}}
\end{figure}
%%%%%%%%%%%%%%%%%%%%%%%%%%%%%%%%%%%%%%%%%%%%%%%%%%%%%
\item $F_V \neq 2 G_V$, $F_A\neq 0$, $(m_V,m_A)<\Lambda$, {\it{i.e.}}, the most generic Higgsless model, containing a single light vector and axial-vector resonance below the cut-off. Due to the large parameter space available, various cancellations can occur (as the ones outlined in the previous two cases), leading to an agreement with EWPT as well as satisfying the unitarity constraint. The various parameter dependences are studied in Fig.~\ref{fig:1c}. 
We first observe that EWPO and the unitarity constraint can only be reconciled for vector masses below $900$~GeV, as shown in the upper-left plot of Fig.~\ref{fig:1c} and confirmed in Fig.~\ref{fig:STU}. Secondly, the upper-right plot in Fig.~\ref{fig:1c} confirms that the axial and vector states should be fairly degenerate, at least for natural values of $F_A$. In the lower plot in Fig.~\ref{fig:1c} we show that for light vector and axial vector states, the model can accomodate the EWPO constraints with reasonable values for the two coupling constants, namely $F_A\lesssim F_V \lesssim 2G_V$, leading to a quite constrained and predictive scenario even in this more general case. 
%%%%%%%%%%%%%%%%%%%%%%%%%%%%%%%%%%%%%%%%%%%%%%%%%%%%%
\end{itemize}
In summary, Fig.~\ref{fig:STU} confirms the general conclusion of Ref.~\cite{Barbieri:2008cc}, namely that accounting for both EWPT and unitarity constraints can be satisfied with at least one light (vector) resonance. However, after including the one-loop resonance contributions to $\hat S$ the allowed parameter ranges are shifted and get significantly more constrained. In particular, two important modifications appear: (1) the generic bound on the lightest vector resonance mass is strengthened to below $1$~TeV and (2) a single vector resonance below the cut-off now requires a sizable $F_V$ coupling, something that is possibly in conflict with bounds from direct searches at the Tevatron.

Generically, in all the studied scenarios the one-loop contributions required to bring $S$ back into the experimental bounds are large and the loop expansion poorly convergent. This is especially troublesome for the stability of the parameters of the model under quantum corrections. However, these potentially large corrections turn out to have a very mild effect for the parameter ranges fitted from $S$ and $T$. The reason can be found in the structure of the corrections to {\it{both}} $S$ and $T$. Let us concentrate on the quadratic pieces, which are the dominant ones. Note that the sum of vector and axial contributions is fixed by the moderate size of $\Delta {\hat{T}}_{UV}$. Since they are small and moreover they enter $\Delta {\hat{S}}_{UV}$ roughly as a difference, the correction to $S$ is actually driven by the $v/m_{V,A}$ pieces coming from the resonance kinetic terms. In other words, the soft breaking of the custodial symmetry is constraining the couplings near the EW scale and thus indirectly protecting the model parameters, in particular $F_V$, from overly large quantum corrections.

\begin{figure}
\begin{center}
  \includegraphics[height=.33\textheight]{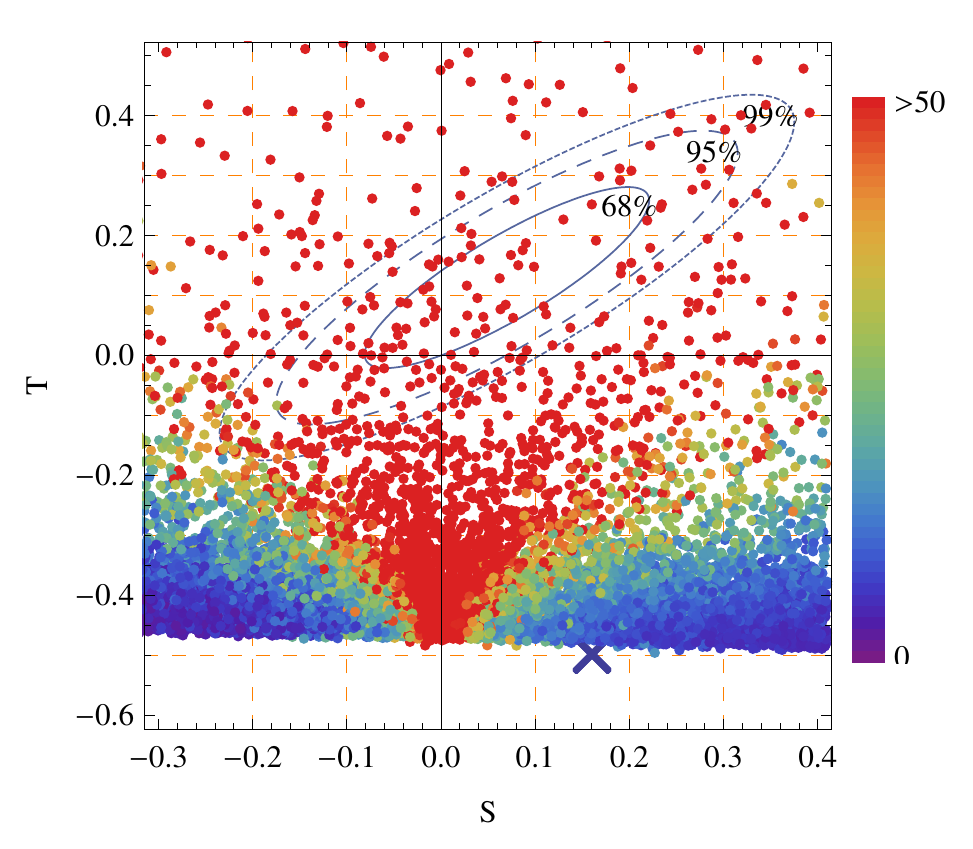}
\end{center}
  \caption{\label{fig:STr} {\it{{\footnotesize{$\hat S$ and $\hat T$ constraints in the generic Higgsless model with a vector and an axial resonance triplet below the cut-off scale $\Lambda=3$~TeV. The reference SM parameters are chosen as in Fig.~1. The parameter space of the model is sampled as discussed in the text. The coloring corresponds to the fine-tuning measure $\Delta$ as defined in the main text.}}}}}
\end{figure}

The increased sensitivity to cancellations among individual contributions to EWPO, as reflected in the very constrained parameter ranges in Figs.~\ref{fig:1a}, \ref{fig:1b} and~\ref{fig:1c} warrants a brief discussion on fine-tuning of the studied scenarios. We evaluate the naturalness of solutions in the parameter space, which satisfy both unitarity and EWPO constraints in the most generic scenario outlined above. We adopt the measure for fine-tuning of ref.~\cite{Barbieri:1987fn} where for each parameter $a_i$ contributing to an observable $\mathcal O_j$ we compute
\begin{equation}
\Delta_{i,j} = \left| \frac{a_i}{\mathcal O_j} \frac{\partial \mathcal O_j(\{a_i\})}{\partial a_i}  \right|\,,
\end{equation}
and then define the maximum fine-tuning of a particular parameter space point as $\Delta = \mathrm{max}\{\Delta_{i,j}\}$. Note that $\Delta=10,50$ corresponds to a $10\%~,2\%$ fine-tuning respectively. We apply $\Delta_{i,j}$ above to $\hat S$ and $\hat T$ and define the relevant parameters as $m_V$, $m_A/m_V$, $G_V \sqrt 3/ v$, $F_V/ 2 G_V$ and $F_A/ F_V$. We scan over this parameter space subject to the UV consistency constraints discussed in Section~2 and to the unitarity constraint on $G_V$ below the cut-off $\Lambda\sim3$~TeV~\cite{Barbieri:2008cc}. The results are shown in Fig.~\ref{fig:STr}.
We observe that the fine-tuning required to be consistent with EWPO at the $68-95\%$ confidence level is at the order or $5-2\%$, respectively. It is also interesting to note that at this level of fine-tuning the 95\% confidence level region with negative $\hat S$ (and $\hat T$) can be reached. However, this region corresponds to the one-loop contributions to $\hat S$ actually overcompensating the tree-level contributions, signaling the breakdown of the loop expansion. 

%%%%%%%%%%%%%%%%%%%%%%%%%%%%%%%%%%%%%%%%%%%%%%%%%%%%
\section{Conclusions}\label{sec:concl}
%%%%%%%%%%%%%%%%%%%%%%%%%%%%%%%%%%%%%%%%%%%%%%%%%%%%

In this paper we have computed the one-loop corrections to the $S$ parameter in the generic Higgsless model considered in Ref.~\cite{Barbieri:2008cc}. Likewise, we have assume the existence of light resonances (a vector and possibly an axial vector state) in the window $v<m_{V,A}<\Lambda\simeq 4\pi v$ whose dynamics are described by a minimal set of operators. This has allowed us to consistently study the predictions of the model for EWPO at next-to-leading order, subject to the unitarity constraint in $W_LW_L$ scattering. 

Our results show that the $S$ and $T$ parameters can be made compatible with the experimental values, but most likely a single vector resonance is not enough. While a light vector is in principle allowed by EWPT and unitarity, direct searches by CDF and D0 require very small values of the coupling $F_V$, something that the fits do not favor. The addition of an axial resonance thus seems to be needed. The picture that eventually emerges turns out to be highly predictive, with the vector mass constrained to the window $500~$GeV$\lesssim m_V\lesssim 1000$~GeV, and the axial one only slightly heavier, $1.2~m_V\leq m_A\leq 1.4~m_V$. The values for the remaining parameters satisfy $F_A\lesssim F_V\lesssim 2G_V\sim 2v/\sqrt{3}$. 

While this is the order of magnitude one would naturally expect for the parameters, one cannot avoid a considerable amount of fine-tuning. The reason, which lies at the heart of Higgsless models, is that the tree level contribution to the $S$ parameter is positive and rather large, and therefore a large and negative one-loop contribution is needed to pull it back to the experimentally allowed region.  In addition, the tree-level contributions proportional to $F_{V,A}/m_{V,A}$ develop quadratic (but not logarithmic) dependences on the cutoff $\Lambda$ at the one-loop order.  In our case, the one-loop contributions scale parametrically as $\Lambda^2/(4\pi m)^2$, $m<1$ TeV being a typical resonance mass and $\Lambda\sim 3$ TeV being the cut-off. We assume the cutoff scale to be close to a physical scale (typically the mass of a heavier vector), such that the estimate of the quadratic corrections is reasonable. In Ref.~\cite{Barbieri:2008cc} the simple addition of two extra operators was shown to do the job quite naturally, at least for $T$. In this paper we have seen that the dynamical mechanism has to be more complicated to also account for $S$: the operators proposed in~\cite{Barbieri:2008cc} not only do not cure the UV sensitivity, but they introduce additional {\it{quartic}} divergences. While they can in general be tamed, the dynamical mechanism loses its main appeal, simplicity: a potentially large number of additional operators might be required. We have however assumed that such mechanism is at work and that the scale $\Lambda$ is representative for higher mass states, not explicitly included as dynamical fields in the model.  It turns out that the soft breaking of custodial symmetry and the moderate corrections to $T$, as preferred by the EWPO fit, actually help to stabilize the parameters of the model against potentially large quantum corrections, leading to a well defined and predictive scenario.

%%%%%%%%%%%%%%%%%%%%%%%%%%%%%%%%%%%%%%%%%%%%%%%%%%%%

%%%%%%%%%%%%%%%%%%%%%%%%%%%%%%%%%%%%%%%%%%%%%%%%%%%%
\section*{Acknowledgments}
%%%%%%%%%%%%%%%%%%%%%%%%%%%%%%%%%%%%%%%%%%%%%%%%%%%%
We thank Gino Isidori for invaluable discussions and initial involvement in this project. This work is supported in part by the EU under contract MTRN-CT-2006-035482 (FLAVIAnet).  O.~C.~is supported by MICINN, Spain under grants FPA2007-60323 and Consolider-Ingenio 2010 CSD2007-00042 –CPAN–. J.~F.~K. is supported in part by the Slovenian Research Agency.
%%%%%%%%%%%%%%%%%%%%%%%%%%%%%%%%%%%%%%%%%%%%%%%%%%%%

\end{document}